\documentclass{jaa}
\usepackage{txfonts}
\usepackage{graphicx}
\usepackage{color}

\usepackage[authoryear]{natbib}

\newcommand{\be}{\begin{equation}}
\newcommand{\ee}{\end{equation}}
\newcommand{\F}{\mathcal{F}}
\newcommand{\n}{\hat{\bf n}}
\def\k{{\bf k}}
   
\def\apj{Astrophysical Journal}
\def\apjl{Astrophysical Journal Letters}

\def\mnras{Monthly Notices of Royal Astronomical Society}

\def\prd{Physical Review D}

\def\k{{\bf k}}

\newcommand{\nh}{{{\rm H}\,{\sc i~}}}
\begin{document}

\title[Cross-correlations of the 21 cm signal with other cosmological probes]
{The redshifted HI 21 cm signal from the post-reionization epoch: Cross-correlations with other cosmological probes}

\author[Guha Sarkar et al.]  {T. Guha Sarkar $^{1}$ \thanks{Email: tapomoy@pilani.bits-pilani.ac.in}, K. K.  Datta$^{2}$ \thanks{e-mail: kanan.physics@presiuniv.ac.in},
 A. K. Pal$^{1}$, T. Roy Choudhury$^{3}$, S. Bharadwaj$^{4}$ 
  \\ $^{1}$Department of
  Physics, Birla Institute of Technology and Science, Pilani 333031,
  India.\\ $^{2}$Department of Physics, Presidency University,
    Kolkata, 700073, India.\\ $^{3}$ National Centre for Radio Astrophysics,
    Pune, 411007, India.\\ $^{4}$Department of Physics, Indian Institute of
    Technology Kharagpur, 721302, India.  }

\maketitle
\date{\today}
\begin{abstract} Tomographic intensity mapping of the \nh  using  the redshifted 21 cm observations opens up 
a new window towards our understanding of cosmological background
evolution and structure formation. This is a key science goal of
several upcoming radio telescopes including the Square Kilometer Array
(SKA). In this article we focus on the post-reionization signal and
investigate the of cross correlating the 21 cm signal with other
tracers of the large scale structure.  We consider the
cross-correlation of the post-reionization 21 cm signal with the
Lyman-$\alpha$ forest, Lyman-break galaxies and late time anisotropies
in the CMBR maps like weak lensing and the Integrated Sachs Wolfe
effect. We study the feasibility of detecting the signal and explore
the possibility of obtaining constraints on cosmological models using
it.
\end{abstract}


\section{Introduction} The tomographic intensity mapping of the neutral hydrogen (\nh) distribution through
redshifted HI 21-cm signal observation is an important 
probe of cosmological evolution and structure formation in 
the post reionization epoch \citep{poreion1, poreion0, poreion4, param2}.  The astrophysical processes dominating  the epoch of
reionization is now  believed to have completed by redshift $z \sim 6$
\citep{fan2006}. In the post-reionization era most of the neutral HI gas
are housed in the  Damped Ly-$\alpha$ (DLA) systems.
These DLA clouds are the predominant source of the HI 21-cm
signal. Intensity mapping involves a low resolution imaging of the diffuse HI 21-cm radiation 
background without attempting to resolve the individual DLAs. Such a tomographic imaging shall naturally provide 
astrophysical and cosmological data  regarding the large scale
matter distribution, structure formation and background cosmic history in the post-reionization epoch \citep{param2, wyithe08,
  param3, camera13, cosmo14}. Several functioning and upcoming radio interferometric arrays
like  Giant Metrewave Radio Telescope
(GMRT) \footnote{http://gmrt.ncra.tifr.res.in/}, the Ooty Wide Field
Array (OWFA) \citep{saiyad2014}, the Canadian Hydrogen Intensity
Mapping Experiment (CHIME)
\footnote{http://chime.phas.ubc.ca/}, the Meer-Karoo Array Telescope (MeerKAT) 
\footnote{http://www.ska.ac.za/meerkat/}, the Square Kilometer Array
(SKA) \footnote{https://www.skatelescope.org/} are 
aimed towards detecting the cosmological 21-cm background radiation.
Detecting the 21 cm signal, is however extremely challenging. This is primarily 
because of the large astrophysical foregrounds \citep{fg1,fg4,fg10} from galactic and extra-galactic sources 
which are several order of magnitude greater than the signal .

Cross-correlating the 21 cm signal with other probes may prove to be useful towards mitigating the
severe effect of foreground contaminants and other systematic effects which
plague the signal.  The main advantage of cross-correlation is that
the cosmological origin of the signal can only be ascertained only if it is
detected with high a statistical significance in the cross-
correlation. Cosmological parameter estimation often involves a joint
analysis of two or more data sets and this would require  not only the
auto-correlation but also cross-correlation information. Further,
the two different  probes may focus on specific $k-$  modes with
high signal to noise ratio and in such cases the cross-correlation
signal takes advantage of the different cosmological  probes simultaneously. This has
been studied extensively in  the case  of the BAO \citep{bharadtapo2} signal.  It is to be noted that
if the observations of the distinct probes are perfect, there shall
be no new advantage of using the cross correlation. However, we expect the
first generation observations  of the redshifted HI 21 cm signal to
have large systematic errors and foreground residuals (even after subtraction).  For a
detection of the 21 cm signal and subsequent cosmological
investigations these measurements can be cross-correlated with other
 large scale structure tracers to yield information from the 21 cm
signal which may not be possible to obtain using  the low SNR auto correlation
signal. In this article we consider the cross-correlation of the 21 cm
signal with the Ly-$\alpha$ flux distribution. On large scales both
the Ly-$\alpha$ forest absorbed flux and the redshifted 21-cm signal
are, believed to be biased tracers of the underlying dark matter (DM)
distribution \citep{mcd03, bagla2, tgs2011, navarro}. The clustering
 of these signals, is then, directly related to the underlying dark
matter power spectrum. We investigate the possibility of using the
cross-correlation of the 21-cm signal and the Ly-$\alpha$ forest for
cosmological parameter estimation, neutrino mass measurement, studying
BAO features and primordial bispectrum.
We also investigate the possibility of correlating the post-reionization 21-cm signal with 
CMBR maps like the weak lensing and ISW anisotropies.

\section{Cross-correlation between cosmological signals (General Formalism) }
Consider two cosmological fields $A(\k)$ and $B(\k)$. These could, for example represent two tracers of large scale structure. We define the cross correlation estimator   $\hat{E}$ as follows
\be
\hat{E} = \frac{1}{2}\  [\ A B^{*} \  + \  B A^{*}\  ]
\ee
We note that  $A$ and $B$ can be  complex fields.
We are interested in the variance 
\be
\sigma_{\hat{E}} ^2 = \langle \  {\hat{E}}^2 \ \rangle  -
{\ \langle \hat{E} \  \rangle  }^2
\ee
Noting that
$
\langle \ A ({\bf{k}})  \ A({\bf{k}}) \  \rangle = \langle \  A({\bf{k}})  \
A^{*}(-{\bf{ k}}) \ \rangle = 0$, we have 
\be
\langle  \ {\hat{E}}^2 \ \rangle = 
\frac{1}{2} \ \left [ \ \langle AA^*\rangle \langle BB^*\rangle \
+ 
\  {|\langle AB \rangle |}^2 \
+
\ 3 \  {|\langle AB^* \rangle |}^2 \ \right ]
\ee
Further, the term $ \langle AB \rangle$ can be dropped
since
\be
\langle \ A ({\bf{k}})  \ B({\bf{k}}) \  \rangle = \langle \  A({\bf{k}})  \
B^{*}(-{\bf{ k}}) \ \rangle = C \delta_{{\bf{k}}, {-{\bf{k}}}}= 0
\ee
This gives
\be
\sigma_{\hat{E}} ^2 = \langle \  {\hat{E}}^2 \ \rangle  -
{\ \langle \hat{E} \  \rangle  }^2 = \ \frac{1}{2}
 [  \ \langle AA^*\rangle \langle BB^*\rangle \ + \
{|\langle AB \rangle |}^2 \ ]
\ee
The variance is suppressed by a factor of $N_c$ for that many number of
independent estimates.
Thus, finally we have
\be
\sigma_{\hat{E}} ^2 = \ = \ \frac{1}{2N_c}
 [  \ \langle AA^*\rangle \langle BB^*\rangle \ + \
{|\langle AB \rangle |}^2 \ ]
\ee

\section{ Cross-correlation of Post-reionization 21 cm  signal with Lyman- $\alpha$ forest}

Neutral gas in the post reionization epoch produces distinct
absorption features, in the spectra of background quasars
\citep{rauch98}.  The Ly-$\alpha$ forest, traces the HI density
fluctuations along one dimensional quasar lines of sight.  The
Ly-$\alpha$ forest observations finds several cosmological
applications \citep{pspec1, Mandel, pspec2, croft1999, cosparam1,
  reion1}.  On large cosmological scales the Ly-$\alpha$ forest and
the redshifted 21-cm signal are, both expected to be biased tracers of
the underlying dark matter (DM) distribution \citep{mcd03, bagla2,
  tgs2011, navarro}. This allows to study their cross clustering
properties in n-point functions.  Also the Baryon Oscillation
Spectroscopic Survey
(BOSS) \footnote{https://www.sdss3.org/surveys/boss.php} is aimed
towards probing the dark energy through measurements of the BAO
signature in Ly-$\alpha$ forest \citep{bossdr11}. The availability of
Ly-$\alpha$ forest spectra with high signal to noise ratio for a large
number of quasars from the BOSS survey allows 3D statistics to be done
with Ly-$\alpha$ forest data \citep{bossdr10, slosar2011}.

Detection these signals are observationally challenging.  For the HI 21-cm
a  detection of the signal requires careful modeling of the
foregrounds \citep{ghosh2011, alonso2014}. Some of the difficulties faced by   Ly-$\alpha$ observations include proper modelling of the continuum,  fluctuations of
the ionizing sources, poor modeling of the temperature-density
relation \citep{mac} and  metal lines contamination in the spectra
\citep{bolton}. The two signals are  tracers of the underlying dark matter distribution. Thus they 
are  correlated on large scales. However foregrounds and
other systematics are uncorrelated between the two independent observations. Hence, the
cosmological nature of a detected signal can be only ascertained in a cross-correlation.
The 2D and 3D cross correlation of the redshifted HI 21-cm
signal with other tracers such as the
Ly-$\alpha$ forest, and the Lyman break galaxies have been proposed as a
way to avoid some of the observational issues \citep{tgs5, navarro2}.
The foregrounds in HI 21-cm observations appear as noise in the cross correlation  and hence, a significant degree
foreground cleaning is still required for a  detection.

We use $\delta_{T}$ to denote the redshifted 21-cm brightness
temperature fluctuations and  $\delta_{\F}$ as the fluctuation in the transmitted flux through the Ly-$\alpha$ forest. 
We write $\delta_{\F}$ and $\delta_{T}$ in Fourier space as
\be
 \delta_{a}({\bf r}) = \int \ \frac{d^3
   {\bf{k}}}{(2\pi)^3} \ e^{i {\bf{k}}. {\bf r}}  \Delta_{a}({\bf{k}}) \,.
\label{eq:deltau}
\ee where $a={\F}$ and $T$ refer to the Ly-$\alpha$ forest transmitted
flux and 21-cm brightness temperature respectively. On large scales we
may write \be \Delta_{a}({\bf{k}}) = C_{a} [1 + \beta_{a} \mu^2]
\Delta({\bf{k}}) \, \ee where $\Delta({\bf{k}})$ is the dark matter
density contrast in Fourier space and $\mu$ denotes the cosine of the
angle between the line of sight direction ${\bf \hat {n}}$ and the
wave vector ($ \mu = {\bf \hat{ k} \cdot \hat{n}}$). $\beta_{a}$ is
similar to the linear redshift distortion parameter.  The corresponding power spectra are 
\be
P_a(k, \mu)=  C_{a}^2 [1 + \beta_{a} \mu^2]^2 P(k)  \ee
where $P(k)$ is the dark matter power spectrum.

For the 21-cm
brightness temperature fluctuations we have \be C_{T} = 4.0 \, {\rm
  {mK}} \, b_{T} \, {\bar{x}_{\rm HI}}(1 + z)^2\left (
\frac{\Omega_{b0} h^2}{0.02} \right ) \left ( \frac{0.7}{h} \right)
\left ( \frac{H_0}{H(z)} \right) \ee The neutral hydrogen fraction
${\bar{x}_{\rm HI}}$ is assumed to be a constant with a value $
      {\bar{x}_{\rm HI}} = 2.45 \times 10^{-2}$ \citep{xhibar, xhibar2, noterdaeme}.  For the HI 21-cm signal the parameter
      $\beta_{T}$, is the ratio of the growth rate of linear
      perturbations $f(z)$ and the HI bias $b_T$. The 21 cm bias is
      assumed to be a consnt. This assumption of linear bias is
      supported by several independent numerical simulations \citep{bagla2, tgs2011} which
      shows that over a wide range of k modes, a constant bias model is
      adequately  describes the 21 cm signal for $z < 3$.  We have adopted  a constant bias $b_{T} = 2$ from
      simulations \citep{bagla2, tgs2011, navarro}.  For the
      Ly-$\alpha$ forest, $\beta_{\F}$, can not be interpreted in the
      usual manner as $\beta_{T}$. This is because Ly-$\alpha$ transmitted flux and the
      underlying dark matter distribution \citep{slosar2011} do not have a simple linear relationship. The
      parameters $(C_{\F}, \beta_{\F})$ are independent of each other.

 We adopt approximately  $(C_{\F}, \beta_{\F}) \approx ( -0.15,
 1.11 )$ from the numerical simulations of Ly-$\alpha$
 forest \citep{mcd03}. We note that for cross-correlation studies the
 Ly-$\alpha$ forest has to be smoothed to the observed frequency resolution of the HI 21 cm
 frequency channels.

We now consider the 3D cross-correlation power spectrum of the HI
21-cm signal and Ly-$\alpha$ forest flux. We consider an observational
survey volume V which on the sky plane  consists of a patch $L \times L$ and
of line of sight thickness $l$ along the radial  direction. We 
consider the flat sky approximation. The Ly-$\alpha$ flux fluctuations
are now written as a 3-D field  \be
\delta_{\F}(\vec{r})= \left[ \ \frac{ {\F}(\vec{r}) -
    {\bar{\F}}}{{\bar{\F}}} \ \right] \ \ee The observed quantity is
$\delta_{{\F} o}(\vec{r}) =\delta_{{\F} o}(\vec{r}) \times
\rho(\vec{r}) $, where the sampling function $\rho(\vec{r})$ is
defined as \be \rho(\vec{r}) =\frac {\sum_a w_a \ \delta_D^{2}(
  \vec{r}_{\perp} - \vec{r}_{\perp a}) }{l \sum_a w_a} \ee and is
normalized to unity ( $\int dV \rho(\vec{r}) = 1$ ).  The summation as
before extends up to $N$. The weights $w_a$ shall in general be
related to the pixel noise.  However, for measurements of
transmitted  hight SNR flux, the effect of the weight functions can
be ignored. With this simplification  we have  used $ w_a =
1$, so that $\sum_a w_a = N$.  In Fourier space we have \be
\Delta_{\F}(\vec{k}) = \int_{-L/2}^{L/2} \int _{-L/2}^{L/2} \int
_{-l/2}^{l/2} d^2{\vec{r_\perp}} dr_{\parallel} \ e ^{ i \vec{k} \cdot
  \vec{r}} \ \delta_{\F}(\vec{r})
\label{eq:ft3}
\ee
One may relate $\vec{k}_\perp $ to $\vec{U}$ as $\vec{k}_\perp= \frac{ 2 \pi \vec{U}}{r}$.
We have,  in  Fourier space  
\be 
 {\Delta}_{{\F} o}(\vec{k}) = \tilde{\rho}(\vec{k}) \otimes
{\Delta_{\F}}(\vec{k}) + \Delta_{N\F}(\vec{k})\ee 
where $\tilde{\rho}$ is the Fourier transform of $\rho$ and  $ \otimes $ denotes a
convolution defined as
\be
\tilde{\rho}(\vec{k}) \otimes
{\Delta_{\F}}(\vec{k}) = \frac{1}{V} \sum_{\vec{k}'}
\ \tilde{\rho}(\vec{k} - \vec{k}') {\Delta_{\F}}(\vec{k}')
\ee
$ \Delta_{N\F}(\vec{k})$ denotes a possible noise term.
Similarly  the 21-cm signal in Fourier space is
 written as
\be
{\Delta}_{{T} o}(\vec{k}) = \Delta_{T}(\vec{k}) +
\Delta_{NT}(\vec{k}) 
\ee
where $\Delta_{NT}$ is the corresponding noise.

 The
cross-correlation 3-D power spectrum $ P_c (\vec{k})$  for the two fields  is defined as
\be
 \langle \ \Delta_{{\F}}(\vec{k}) \Delta^{*}_{T}(\vec{k}') \  \rangle = V  P_c (\vec{k}) \delta_{\vec{k}, \vec{k}'}
\label{eq:fluxps3}
\ee
Similarly,  we define the two auto-correlation multi frequency
angular power spectra, $ P_T (\vec{k})$ for 21-cm radiation
and $P_{\F}(\vec{k})$ for Lyman-$\alpha $ forest flux
fluctuations as
\be
 \langle \ \Delta_{T}(\vec{k}) \Delta^{*}_{T}(\vec{k}') \  \rangle = V  P_T (\vec{k}) \delta_{\vec{k}, \vec{k}'}
\ee
\be
 \langle \ \Delta_{\F}(\vec{k}) \Delta^{*}_{\F}(\vec{k}') \  \rangle = V  P_{\F} (\vec{k}) \delta_{\vec{k}, \vec{k}'}
\ee
 We define the cross-correlation estimator  $\hat{\mathcal{E}}$ as
\be
\hat{\mathcal{E}}( \vec{k}, \vec{k}' ) = \frac{1}{2} \left[  {\Delta}_{{\F} o}(\vec{k})
{\Delta}^{*}_{{T} o}(\vec{k}') +  {\Delta}^{*}_{{\F}  o}(\vec{k}) {\Delta}_{{T} o}(\vec{k}') \right]
\ee
We are interested in the various statistical properties of this estimator.
Using the definitions of ${\Delta}_{{\F} o}(\vec{k})$ and
${\Delta}_{T  o}(\vec{k})$ we have the expectation value of
$\hat{\mathcal{E}}$ as
\begin{eqnarray}
\langle \  \hat{\mathcal{E}}( \vec{k}, \vec{k}' ) \ \rangle = 
\frac{1}{2} 
\langle \ [\tilde{\rho}(\vec{k}) \otimes
{\Delta_{\F}}(\vec{k}) + \Delta_{N\F}(\vec{k})] 
\times [ \Delta^{*}_{T}(\vec{k}') +
\Delta^{*}_{NT}(\vec{k}')] \ \rangle \nonumber  \\ 
+ \frac{1}{2} 
\langle \ [\tilde{\rho}^{*}(\vec{k}) \otimes
{\Delta_{\F}}^{*}(\vec{k}) + \Delta_{N\F}^{*}(\vec{k})]
\times [\Delta_{T}(\vec{k}') +
\Delta_{NT}(\vec{k}')  ] \ \rangle 
\end{eqnarray}
We  assume that the quasars are distributed in a random fashion,  are
not clustered and  the different noises are uncorrelated. Further, we
note that the
quasars are assumed to be at a redshift different from rest of the
quantities and hence $\rho$ is  uncorrelated with
both $\Delta_T$ and $\Delta_{\F}$.
Therefore
we have
\be
\langle \  \hat{\mathcal{E}}( \vec{k}, \vec{k}') \ \rangle = \frac{1}{V} 
\sum_{\vec{k''}} \langle  \tilde{\rho} ( \vec{k} - \vec{k}'') \rangle
\times V  P_{\F T} (\vec{k}'')   \delta_{\vec{k}'', \vec{k}'}
\ee
Noting that 
\be
\langle \tilde{\rho}(\vec{k}) \rangle = \delta_{ \vec{k}_{\perp},0}\delta_{ \vec{k}_{\parallel},0}
\ee
we have
\be
\langle \  \hat{\mathcal{E}}( \vec{k},\vec{k}' ) \ \rangle = P_{\F T} (\vec{k}) \delta_{\vec{k}, \vec{k}'} 
\ee
Thus,  the expectation value of the estimator faithfully returns the
quantity we are probing, namely the 3-D cross-correlation power spectrum $ P_{\F T} (\vec{k})$.

We next consider the variance of the estimator $\hat{\mathcal{E}}$ defined as
\be
\sigma_{\hat{\mathcal{E}}}^2 = \langle {\hat{\mathcal{E}}}^2 \rangle - {\langle \hat{\mathcal{E}}
  \rangle }^2
\ee
\begin{eqnarray}
\sigma_{\hat{\mathcal{E}}}^2 = \frac{1}{2} \langle  \Delta_{{\F} o} (\vec{k})
\Delta_{{\F} o} ^{*}(\vec{k})  \rangle  \langle \Delta_{T o}
(\vec{k}') \Delta_{T o}^{*}(\vec{k}') \rangle 
+  \frac{1}{2} {| \langle \Delta_{{\F} o} (\vec{k}) \Delta_{T o}^{*}
(\vec{k}')    \rangle |}^2
\end{eqnarray}
We saw that 
\be
\langle \Delta_{{\F} o} (\vec{k}) \Delta_{T o}^{*}
(\vec{k}')\rangle  =  P_{\F T} (\vec{k})  \delta_{\vec{k}, \vec{k}'} 
\ee
and we note that
\be
 \langle \Delta_{T o}
(\vec{k}) \Delta_{T o}^{*}
(\vec{k}) \rangle = V [ P_T ( \vec{k}) + N_T (\vec{k})]
\ee
where $N_T$ is the 21-cm noise power spectrum. We also have for the Ly-$\alpha$ forest 
\be
 \langle  \Delta_{{\F} o} (\vec{k})
\Delta_{{\F} o} ^{*}(\vec{k})  \rangle = 
\langle \  \tilde{\rho}(\vec{k}) \otimes
{\Delta_{\F}}(\vec{k})  \  \tilde{\rho}^{*}(\vec{k}) \otimes
{\Delta_{\F}}^{*}(\vec{k}) \  \rangle
+ N_{\F} L^2
\ee
where $ N_{\F}$ is the Noise power spectrum corresponding to the
Ly-$\alpha$ flux fluctuations.  Using the relation 
\be
\langle \ \tilde{\rho}(\vec{k}) \tilde{\rho}^{*}(\vec{k'})\ \rangle =
\frac{1}{N} \delta _{\vec{k_\perp}, \ \vec{k'_\perp}} \delta_{k_{\parallel,0}}  \delta_{k'_{\parallel,0}}  + (1 - \frac{1}{N}) \delta_{\vec{k}, 0
} \delta_{\vec{k'}, 0}
\ee
we have
\begin{eqnarray}
 \langle  \Delta_{{\F} o} (\vec{k})
\Delta_{{\F} o} ^{*}(\vec{k})  \rangle  = 
\frac{1}{V^2}  \sum_{\vec{k_1},\vec{k_2}} \langle \tilde{\rho}(\vec{k} -\vec{k}_1 )  \tilde{\rho}^{*}(\vec{k} - \vec{k}_2 )\rangle 
\langle \Delta_{\F}(\vec{k}_{1\perp}, k_{1\parallel}) \Delta_{\F}^{*}(\vec{k}_{2\perp}, k_{2\parallel})\rangle
\end{eqnarray}
or
\begin{eqnarray}
 \langle  \Delta_{\mathcal{F}o} (\vec{k}) \Delta_{\mathcal{F}o} ^{*}(\vec{k})  \rangle
 = \frac{1}{V^2}  \sum_{\vec{k_1},\vec{k_2}} 
 \delta_{(\vec{k}-\vec{k_1}),0} \delta_{(\vec{k}-\vec{k_2}),0}
 \nonumber 
+ \frac{1}{N} \left (  \delta_{(\vec{k_\perp}-\vec{k_{1\perp}}),(\vec{k_\perp}-\vec{k_{2\perp}}) }\delta_{({k_\parallel}-{k_{1\parallel}}),({k_\parallel}-{k_{2\parallel}}) }\right) \nonumber \\
\times \langle \Delta_{\mathcal{F}}(\vec{k}_{1\perp}, k_{1\parallel}) \Delta_{\mathcal{F}}^{*}(\vec{k}_{2\perp}, k_{2\parallel})\rangle 
\end{eqnarray}

This gives
\begin{eqnarray}
\sigma_{\hat{\mathcal{E}}}^2 = \frac{1}{2} \left [ \frac{1}{N} \sum_{\vec{k_\perp}} P_{\F}( \vec{k})  +  P_{\F}( \vec{k})
+ N_{\F} \right ] \times  \left [ P_T ( \vec{k}) + N_T \right ]  
+  \frac{1}{2} P_{\F T}^2
\end{eqnarray}

Writing the summation as an integral we get
\begin{eqnarray}
\sigma_{\hat{\mathcal{E}}}^2 = \frac{1}{2} \left [ \frac{1}{\bar{n}} \left ( \int d^2 \vec{k_\perp} \  P_{\F}( \vec{k}) \right) +  P_{\F}( \vec{k})
+ N_{\F} \right ] 
\times  \left [ P_T ( \vec{k}) +  N_T \right ] \nonumber  
 +  \frac{1}{2} P_{\F T}^2
\label{eq:sigma3}
\end{eqnarray}
where
$\bar{n}$ is the angular density of quasars and $\bar{n} = N/ L^2$.
We assume that the variance $\sigma^2_{{\F} N}$ of  the pixel noise contribution
to $\delta_{\F}$ is a constant and is  same across all the quasar spectra  whereby we
have $N_{\F} = {\sigma}^2_{{\F} N}/\bar{n}$ for its noise power spectrum. 
 An uniform weighing scheme for all quasars is a good approximation  when most of the spectra are measured with a sufficiently
high SNR \citep{mcquinnwhite}.
We have not incorporated  quasar clustering which is supposed to be sub-dominant  as compared to Poisson noise. In
reality, the clustering would enhance the term $ \left (P^{\rm
  {1D}}_{\F \F}(k_{\parallel}) P^{\rm {2D}}_{w} + N^{}_{\F} \right )$
  by a factor $ \left ( 1 + {\bar n}
  C_{Q}({\bf k }_{\perp}) \right) $, where $C_{Q}({\bf k }_{\perp})$
  is the angular power spectrum of the quasars\citep{myers1}. 

For  a radio-interferometric measurement of the 21-cm signal
we have \citep{fg2, param1} 
\be
N_{T} (k,\nu)=\frac{T_{sys}^2}{B t_0} \left ( \frac{\lambda^2}{A_e}
\right)^2 \frac{r_{\nu}^2 L}{n_b(U,\nu)} 
\label{eq:noise-error}
\ee 
Here $T_{sys}$ denotes the system temperature. $B$is  the observation bandwidth,  $t_0$ is the total 
observation time, $r_{\nu}$ is the comoving distance to the redshift
$z$ , $n_b(U,\nu)$ is the density of baseline $ U$, and $A_e$ is the effective collecting area of each  antenna.

\subsection{The cross correlation signal and constraints with SKA}

\begin{figure}
\begin{center}
\includegraphics[height=6cm, width=6cm, angle=-90]{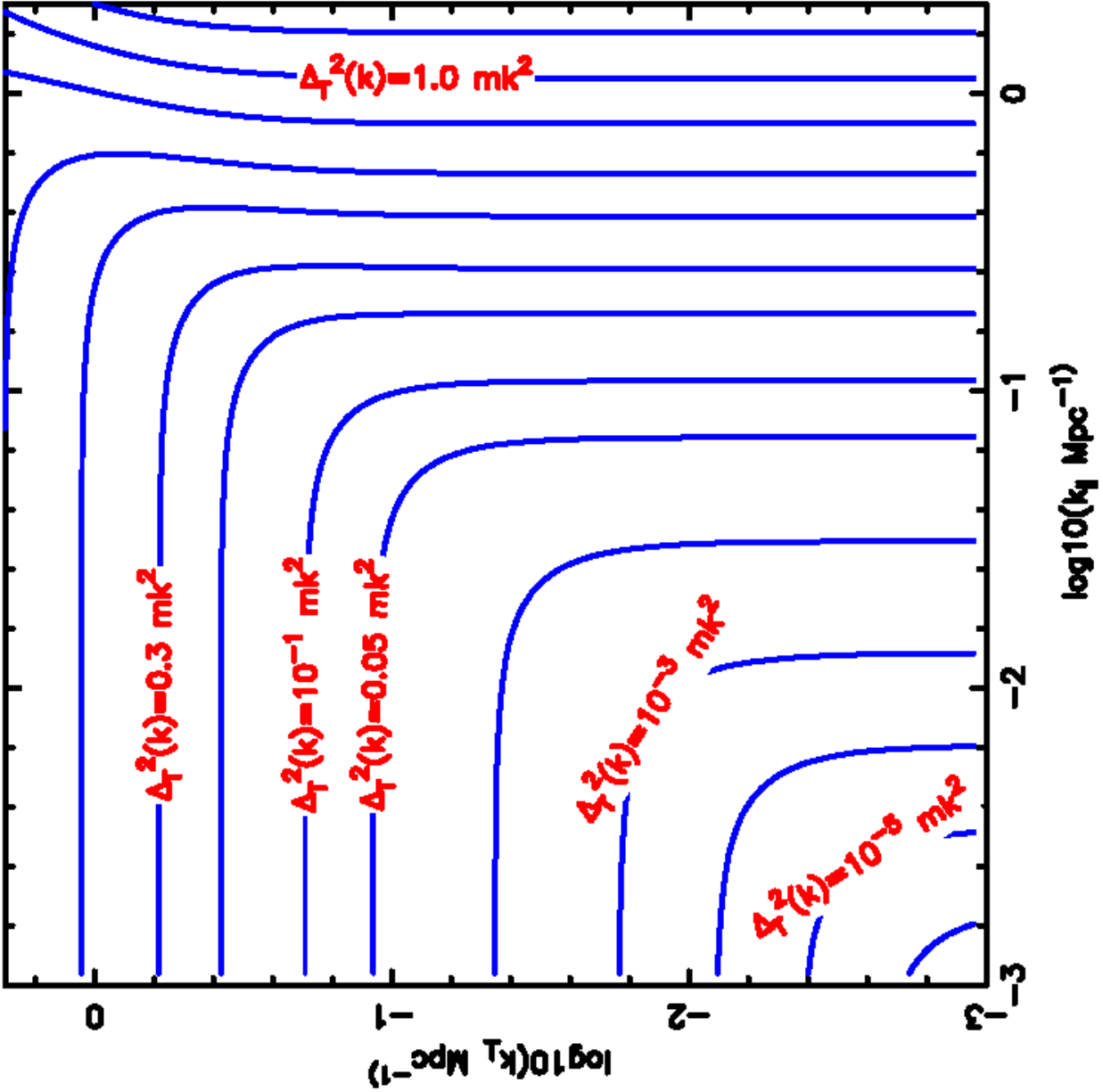}
\includegraphics[height=6cm, width=6cm, angle=-90]{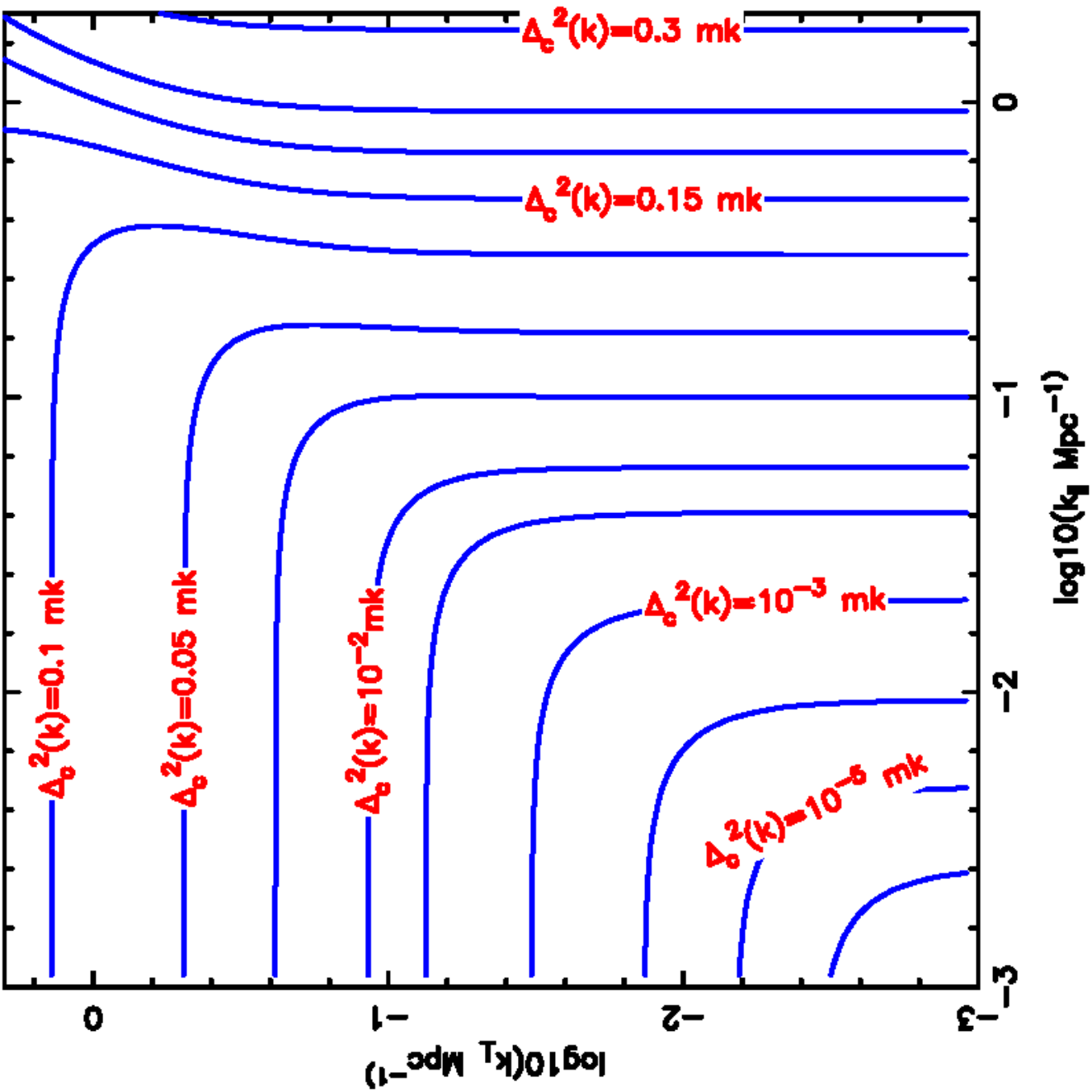}
\end{center}
\caption{Figure showing the  power spectrum in 3D redshift space at  $z = 2.5$. The left panel shows the HI 21-cm  power spectrum 
$\Delta_T^2 = k^3 P_{TT}({\bf k})/ 2 \pi^2$ and the right panel shows the 3D
cross-correlation power spectrum $ \Delta_C^2 =  k^3 P_{T \F}({\bf k})/ 2 \pi^2$. The  redshift space distortion reveals as departure from spherical symmetry of the power 
spectrum. (\cite{TGS15})}
\label{fig:sigHips}
\end{figure}

We investigate the possibility of detecting the signal using the
upcoming SKA-mid phase1 telescope and future Ly-$\alpha$ forest
surveys with very high quasar number densities. Two separate
telescopes named SKA-low and SKA-mid operating at two different
frequency bands and will be constructed in Australia and South Africa
respectively in two phases.  For this work we consider the instruments
SKA1-mid which will be built in phase 1. The instrument specifications
such as the total number of antennae, antenna distribution, frequency
coverage, total collecting area etc., have not been fixed yet and
might change in future. We use the specifications considered in the `Baseline Design Document' and 'SKA Level
1 Requirements (revision 6)' which are available on the SKA
website\footnote{https://www.skatelescope.org/key-documents/}.  We
assume that the SKA1-mid will operate in the frequency range from $350
\,{\rm MHz}$ to $14$ GHz. It shall have  $250$ antennae of
$7.5$ meters radius each.  We use the baseline distribution given in \citep{navkkd} 
 (figure 6-blue line) for the calculation presented
here. We note that, the baseline distribution used here is consistent
with the projected antenna layout distribution with $40 \%$, $54 \%$,
$70 \%$, $81 \%$ and $100 \%$ of the total antennae are assumed to be  enclosed
within $0.4$ km, $1$ km, $2.5$ km, $4$ km and $100$ km radius
respectively.

The fiducial redshift of $z=2.5$ is justified since the quasar distribution
 peaks in the  range $ 2 < z < 3$.  Only a smaller part of the quasar spectra corresponding to an 
approximate band $\Delta z
\sim 0.4$ is used  to avoid contamination from metal lines and  quasar proximity effect. 
The cross-correlation can however only be computed in the
region of overlap between the 21-cm signal and the Ly-$\alpha$ forest
field.

The left panel of the figure (\ref{fig:sigHips}) shows the dimensionless redshift space 21-cm power spectrum
($\Delta^2_T(k_{\perp}, k_{||})=k^3 P_T(k_{\perp}, k_{||})/2 \pi^2$)
at $z = 2.5$.  We
  can see that the power spectrum is not circularly symmetric in the  $(k_{||}$,  $k_{\perp})$ plane. The
  asymmetry is related to the  redshift space distortion
  parameter.  The right panel of figure (\ref{fig:sigHips})
  shows the 21-cm and Ly-$\alpha$ cross-power   spectrum.
  \begin{figure}
\begin{center}
\includegraphics[height= 6cm, width= 6cm, angle=-90]{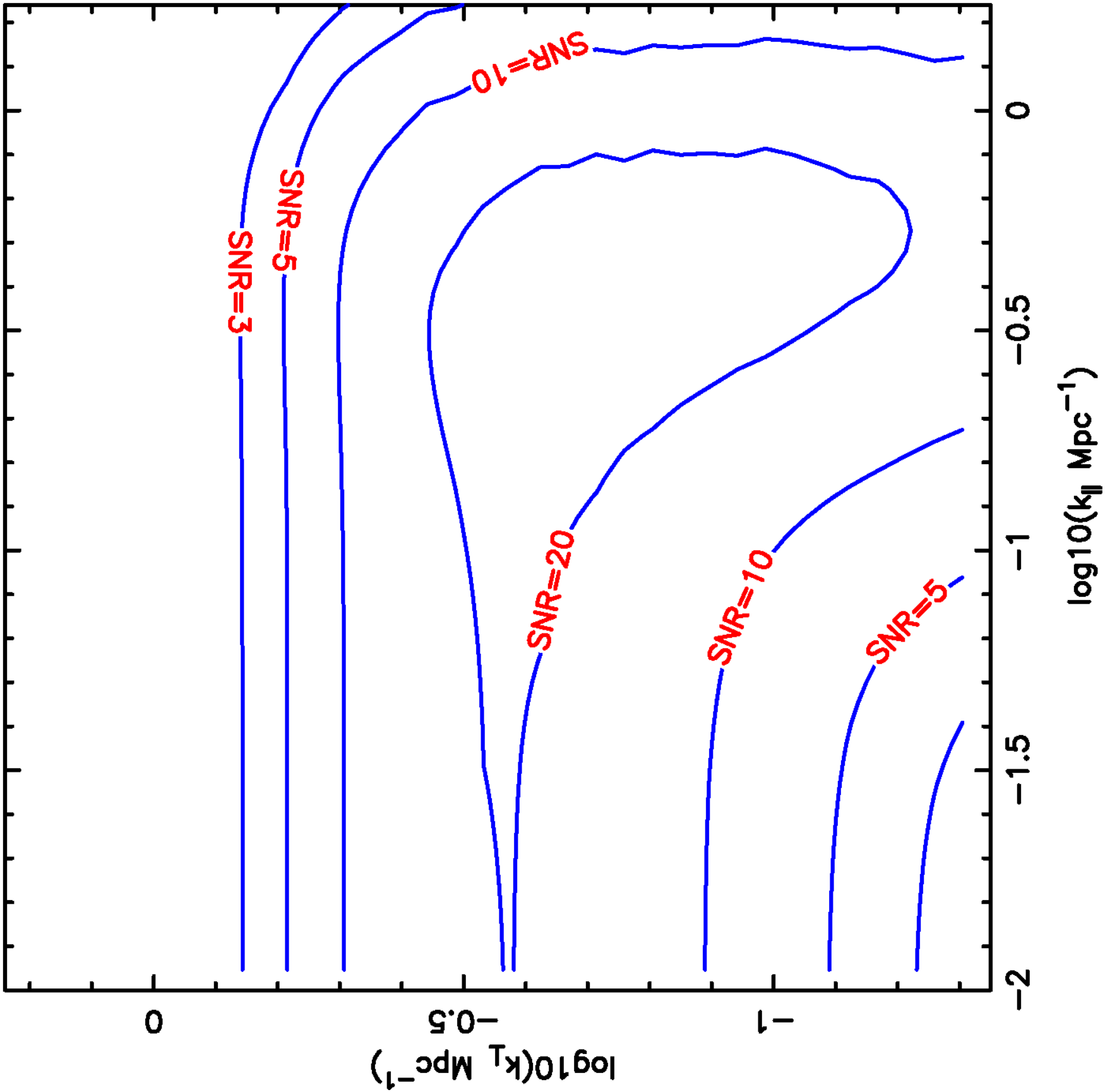}
\includegraphics[height= 6cm, width= 6cm, angle=-90]{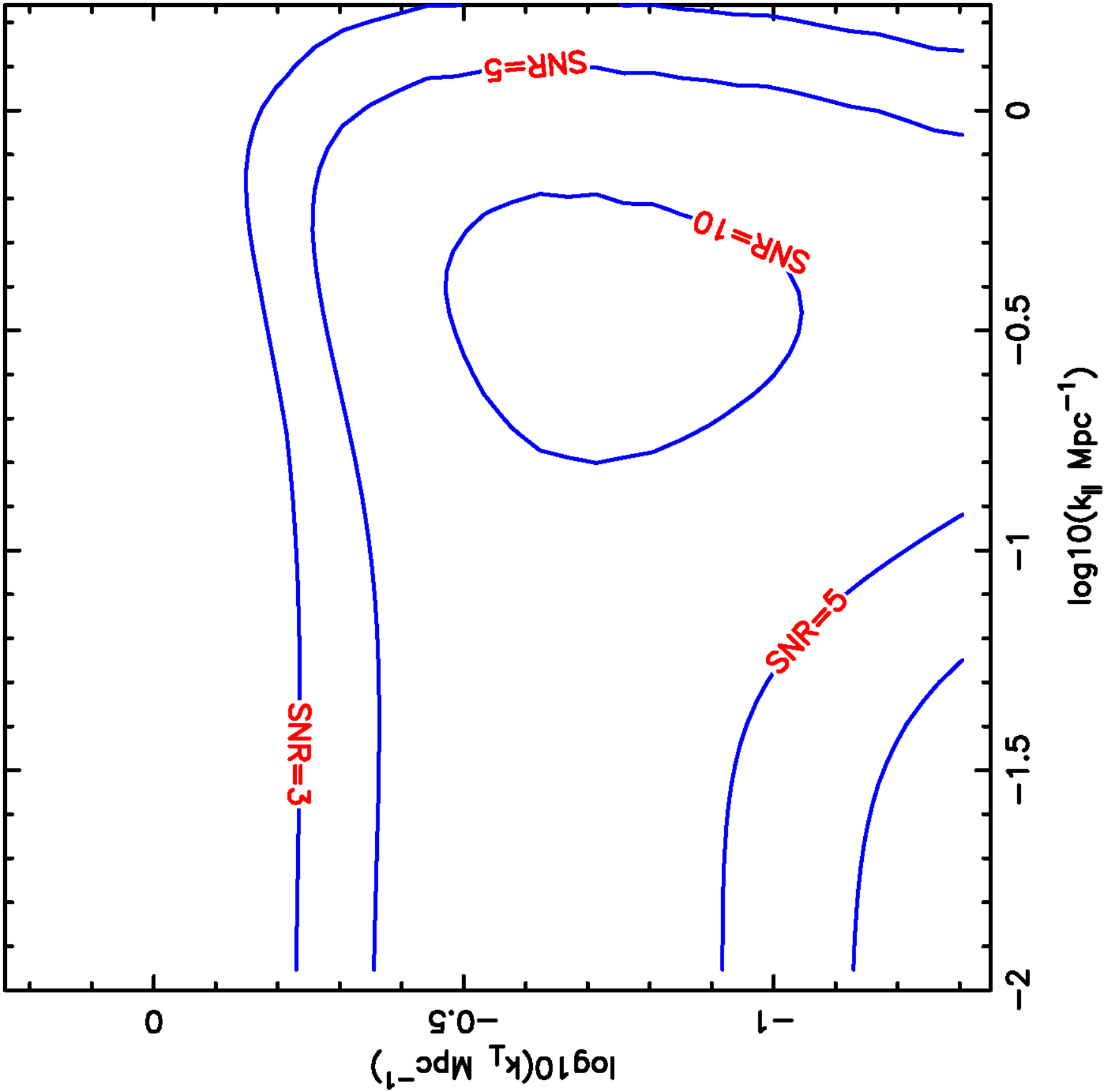}
\end{center}
\caption{The left panel shows SNRcontours for the 21-cm auto-correlation power
spectrum in redshift space at z = 2.5. We have considered a 400 hrs
observation at 405 MHz and assumed that complete foreground cleaning
is done. The right panel shows the SNR contours for the
cross-correlation signal (\cite{TGS15})}.
\label{fig:Hipps-nofg}
\end{figure}

We first consider that a perfect foreground 
subtraction is achieved. The left panel of the figure
(\ref{fig:Hipps-nofg}) shows the contours of SNR for the 21-cm auto
correlation power spectrum for a $400 \rm hrs$ observation and total
$32$MHz bandwidth at a frequency $405.7 \rm MHz$. We have taken a bin 
$(\Delta k, \Delta \theta)=(k/5, \pi/10)$. The SNR reaches at the peak
($>20$)at intermediate value of $(k_{\perp}, k_{\parallel})=(0.4,0.4)
\, {\rm Mpc^{-1}}$ . We find that $5\sigma$ detection is possible in the range $0.08
\lesssim k_{\perp} \lesssim 0.6 \, {\rm Mpc^{-1}}$ and $0.1 \lesssim
k_{\parallel} \lesssim 1.5 \, {\rm Mpc^{-1}}$. The  range for
the $10\sigma$ detection is $0.12 \lesssim k_{\perp} \lesssim 0.5 \,
{\rm Mpc^{-1}}$ and $0.2 \lesssim k_{\parallel} \lesssim 1.2 \, {\rm
  Mpc^{-1}}$. At lower values of $k$ the noise is expected to be  dominated by cosmic
variance whereas, the noise is predominantly of instrumental origin  at large
$k$.

The right panel of the figure (\ref{fig:Hipps-nofg}) shows the SNR contours
 for the Ly-$\alpha$ 21-cm cross-correlation power
spectrum. For the 21 cm signal, a $400 \rm hrs$ observation is
considered. We have taken  $\bar n = 30
\rm deg^{-2}$, and the Ly-$\alpha$ spectra are assumed to be measured
at a $2\sigma$ sensitivity level. We use $\beta_F$ to be $1.11$ and
overall normalization factor $C_{\F}=-0.15$ consistent with recent
measurements \citep{slosar2}.  Although the overall SNR for the cross
power spectrum is lower compared to the 21-cm auto
power spectrum, $5\sigma$ detection is ideally possible for the
$0.1 \lesssim k_{\perp} \lesssim 0.4 \, {\rm Mpc^{-1}}$ and $0.1
\lesssim k_{\parallel} \lesssim 1 \, {\rm Mpc^{-1}}$. The SNR peaks ($>10$)at
$(k_{\perp}, k_{\parallel})\sim (0.2,0.3) \, {\rm Mpc^{-1}}$.
 The error in the cross-correlation can be reduced either by increasing the QSO
number density or by increasing the observing time for HI 21-cm survey. The QSO
number density is already in the higher side for the BOSS survey that we consider. The only  way to reduce the
variance is to consider more observation time for HI 21-cm survey and enhance the volume of the survey.

\subsection{Parameter estimation using the cross-correlation}
\begin{figure}
\begin{center}
\includegraphics[width=14cm, angle=0]{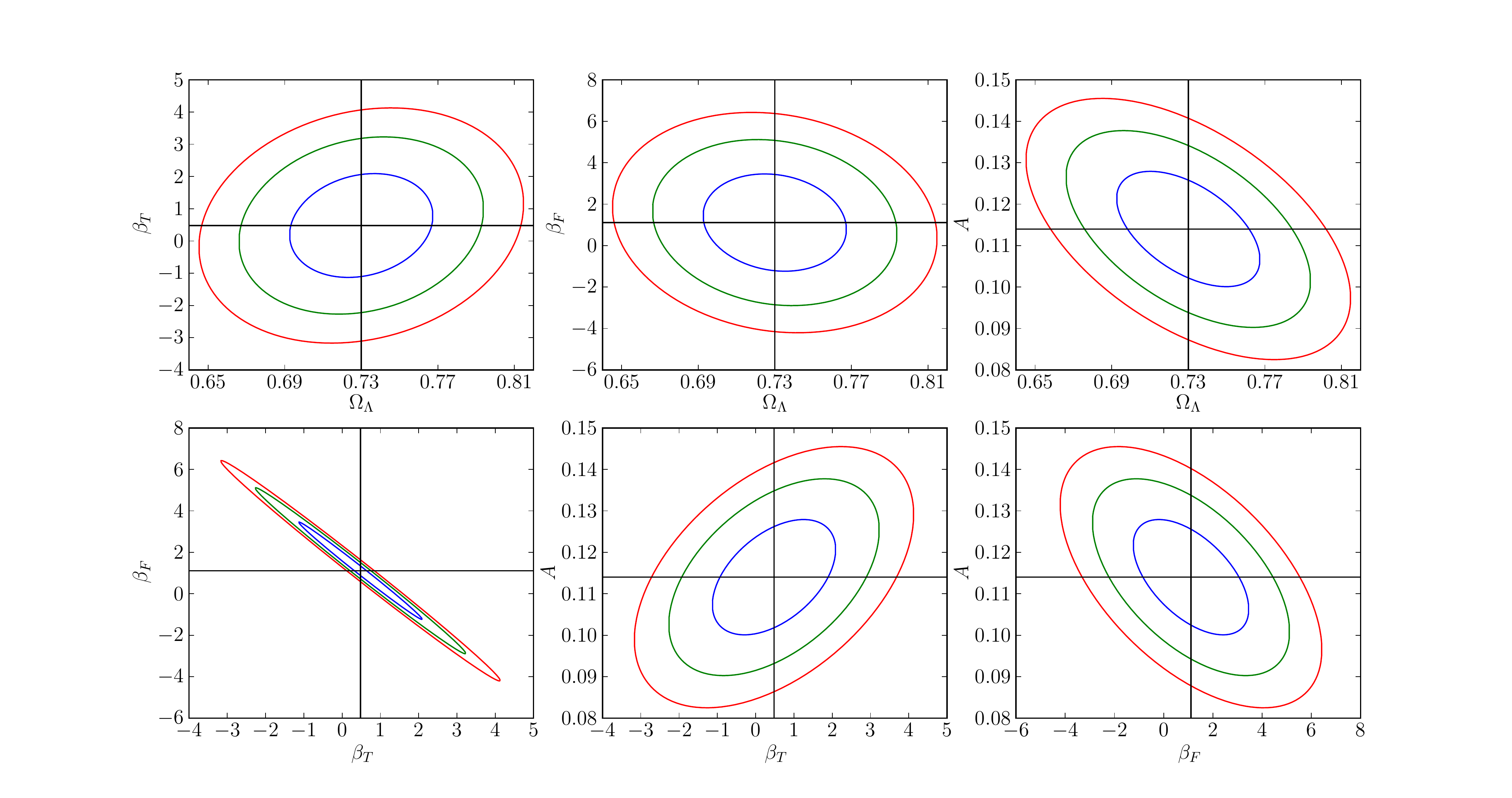}
\caption{The $68.3 \%, 95.4 \%$ and $99.8\%$ confidence
  ellipses for the parameters $(A, \beta_{T}, \beta_{\F},  \Omega_{\Lambda})$. (\cite{TGS15})}
\label{fig:ellipse}
\end{center}
\end{figure}
We now consider the precision at which we can constrain various model
parameters using the Fisher matrix analysis.  Figure
(\ref{fig:ellipse}) shows the $68.3 \%, 95.4 \%$ and $99.8\%$
confidence contours obtained using the Fisher matrix analysis for the
parameters $(A, \beta_{T}, \beta_{\F}, \Omega_{\Lambda})$. The table
\ref{tab:ska1-low} summarises the $1-\sigma$ error these
parameters. The parameters $(\Omega_{\Lambda}, A)$ are constrained
much better that $\beta_{\F}$ and $\beta_{T}$ at $(3.5 \%, 8 \%)$. The error projections presented here are for a single field of view radio observation.
 The noise scales as $ \sigma/
\sqrt N$ where $N$ is the number of pointings. 
\begin{table}
\begin{center}
\caption{This shows $1-\sigma$ error on various cosmological parameters for a single field observation.} 
\centering 
\vspace{.2in}
\begin{tabular}{crrr} 
\hline
\hline 
 Parameters  & Fiducial Value & $1 \sigma$ Error &  $1 \sigma$ Error\\ 
&&(marginalized)&(conditional)\\
\hline
$\beta_{T}$ & 0.48 & 1.06 & 0.04 \\
\hline
$\beta_{\F}$ & 1.11 & 1.55 & 0.05 \\
\hline
$\Omega_{\Lambda}$ & 0.73 & 0.025 & 0.013\\
\hline
$A$ & 0.114 & 0.01 & 0.002\\
\hline
\end{tabular}
\label{tab:ska1-low}
\end{center}
\end{table}

 We also consider conditional error on each of the parameters assuming
 that the other three are known.  The projected $1-\sigma$ error in $\beta_T$ and
 $\beta_{\F}$ are $8.5 \%$ and $4.5\%$ respectively for single
 pointing.  For $10$ independent radio observations
 the conditional errors improve to $2.7 \%$, $1.4\%$, $0.4\%$ and
 $0.6\%$ for $\beta_T$, $\beta_{\F}$, $\Omega_{\Lambda}$ and $A$
 respectively.  These constraints on the redshift space
 distortion parameters $\beta$ from our cross-correlation analysis 
are found to be quite  competitive with other cosmological
 probes \citep{andreu2,slosar2011}.  Further, we note that higher density
 of QSOs and improved SNR for the individual QSO spectra shall also
provide stronger  constraints.

\subsection{BAO imprint on the cross-correlation signal }
The characteristic scale of the BAO is set by the acoustic  horizon $s$ at
the epoch of recombination The comoving length-scale $s$ defines a
angular scale $ \theta_s = s [(1+z) D_A(z)]^{-1}$ in the transverse direction and a
radial redshift interval $\Delta z_s =s H (z)/c$, where $D_A(z)$ and
$H(z)$ are the angular diameter distance and Hubble parameter
respectively.  The comoving acousic horizon scale $s= 143 \, \rm Mpc$
correspond  to  an angle $\theta_s= 1.38^{\circ}$ and reshift interval $\Delta z_s = 0.07$ at redshift $z
= 2.5$.  Measurement of $\theta_s$ and $\Delta z_s$ separately, allows
the determination of $ D_A(z)$ and $H(z)$ separately and thereby
constrain background cosmological evolution.  Here we consider the
possibility of measurement of these two parameters from the imprint of
BAO features on the cross-correlation power spectrum.

The Fisher matrix is given by \citep{bharadtapo2}
\be 
F_{i j}  = \frac{V}{(2 \pi)^3} \int \,  \, 
\frac{d^3  \k}{[P^2_{\F T}(\k)+ P_{\F \F o}(\k) P_{TTo}(\k)]}
 \frac{\partial P_{\F T}(\k)}{\partial q_i}
 \frac{\partial P_{\F T}(\k)}{\partial q_j}
\label{eq:pe3}
\ee

where $q_i$ refer to the cosmological parameters to be constrained.
This BAO signal  is mainly present at small ${\bf k}$ (large
scales) with the first peak at roughly $ k \sim 0.045 {\rm
  Mpc}^{-1}$. The subsequent oscillations are highly  suppressed by $
k \sim 0.3 {\rm Mpc}^{-1}$ which is within the limits of the
$\k_{\perp}$ and $k_{\parallel}$ integrals. We use $P_b=P-P_c$ to
isolate the purely baryonic features, and we use
this in  $\partial P(k)/\partial q_i$.  Here, $P_c$
is the  CDM power spectrum without the baryonic features.  This
gives \be P_b(\k) = \sqrt{8 \pi^2} A \, \frac{\sin x}{x} \, \exp
\left[ - {\left(\frac{k}{k_{silk}} \right ) }^{1.4} \right] \, \exp
\left [{ - \left( \frac{k^2}{2 k_{nl}^2}\right ) }\right] \ee where
${k_{silk}}$ and ${k_{nl}}$ denotes the scale of `Silk-damping' and
`non-linearity' respectively.  We have used $k_{nl} = (3.07 \, h^{-1}
\rm Mpc)^{-1}$ and $k_{silk} = (7.76 \, h^{-1}\rm Mpc)^{-1}$ from
\citep{seoeisen07}. The quantity $ x = \sqrt { k_{\perp}^2 s_{\perp}^2
  + k_{\parallel}^2 s_{\parallel}^2}$ where $ s_{\perp}$ and $
s_{\parallel}$ corresponds to $\theta_s$ and $\Delta z_s$ in units of
distance. $A$ is an overall normalization constant. The value of $s$ is
well constrained from  CMBR data. Changes in $D_A$ and $H(z)$ manifest as the corresponding 
changes in the values of $ s_{\perp}$ and $
s_{\parallel}$ respectively, and thus the fractional errors in $ s_{\perp}$
and $ s_{\parallel}$ correspond to fractional errors in $D_A$ and
$H(z)$ respectively.  We choose $q_1 = {\rm ln} (s_{\perp}^{-1})$ and
$q_2 = {\rm ln}( s_{\parallel})$ as the cosmological parameters to be constrained, and
determine the precision at which it will be possible to measure
these using the BAO imprint on the  in the cross-correlation power spectrum.  We use the formalism outlined in \citep{seoeisen07}, whereby we construct the $2-D$ Fisher matrix
\be F_{ij} = V A^2 \int d k \int_{-1}^{1} d \mu \frac{k^2 \exp[-2
    (k/k_{silk})^{1.4} - (k/k_{nl})^2]} {[P_{\F T}^2(k)+ P_{\F \F
      o}(\k) P_{TTo}(\k)/F^2_{\F T}(\mu)]} f_i(\mu) f_j(\mu)
\label{eq:fmf}
 \ee 
\be
F_{\F T}(\mu)= \frac{H(z)}{r^2 c} C_{\F} C_{T} \, (1 + 
\beta_{\F} \mu^2)  (1 + \beta_{T} \mu^2)
\label{eq:cc6}
\ee
where  $f_1= \mu^2 -1$ and  $f_2= \mu^2$. The Cramer-Rao bound 
$\delta q_i = \sqrt{F^{-1}_{ii}}$ is used to calculate the maximum theoretical error in the parameter $q_i$.
 A combined distance measure $D_V$, also referred to as the
  ``dilation factor''  \citep{baoeisen-05} 
\be
D_V(z)^3 = (1 + z) ^2 D_A(z) \frac{c z}{H(z)}
\ee
is often used as a single  parameter to quantify BAO observations .
We use $\delta D_V/D_V = \frac{1}{3} ( 4 F^{-1}_{11} +
 4 F^{-1}_{12} +  F^{-1}_{22} )^{0.5}$ to obtain the relative 
 error in $D_V$. The dilation factor is known to be  particularly 
 useful when the  individual measurements of $D_A$
 and $H(z)$ have low signal to noise ratio.

 The Fisher matrix formalism is used to determine the accuracy with
 which it will be possible to measure cosmological distances using
 this cross-correlation signal.

 The limits $\bar{n}_Q \rightarrow \infty$ and
$N_T \rightarrow 0$, which correspond to $P_{\F \F o} \rightarrow
P_{\F \F}$  and $P_{T T o} \rightarrow P_{T T}$, set the 
  cosmic variance limit.
  In this limit,  where the SNR depends only on the survey volume corresponding to the
 total field of view we have $\delta
  D_V/D_{V} = 0.15 \%$,  $\delta H/H = 0.25 \%$  and   $D_A/D_{A} = 0.15
  \%$ which are independent of
  any   of the other  observational details.  The fractional errors
  decrease slowly  beyond $\bar{n}_{Q} > 50 \rm
  deg^{-2}$ or $N_T < 10^{-6} \rm mK^2$. 
We find that  parameter values $\bar{n}_{Q} \sim 6 \, \rm deg^{-2}$ and $N_T
\sim 4.7 \times 10^{-5} \ \rm mK^2$, attainable with BOSS and SKA1 mid are adequate for a $1 \%$
accuracy, whereas $\bar{n}_{Q} \sim 2 \, \rm deg^{-2}$ and $N_T \sim
3 \times 10^{-3} \ \rm mK^2$ 
are adequate for a $\sim 10 \%$ accuracy in measurement of $D_V$. 
 With a BOSS like survey  is possible to achieve the fiducial value 
$\delta D_V/D_V= 2.0 \%$ from the cross-correlation at $N_T = \, 2.9
\, \times \, 10^{-4}{\rm mK}^2$.  The error varies slower than
$\sqrt{N_T}$ in the range $N_T = 10^{-4}{\rm mK}^2$ to $N_T =
10^{-5}{\rm mK}^2$. We have $(\delta D_V/D_V,\delta D_A/D_A,\delta
H/H)=(1.3, 1.5, 1.3) \, \%$ and $(0.67, 0.78, 0.74) \, \%$ at $N_T =
10^{-4}{\rm mK}^2$ and at $N_T = 10^{-5}{\rm mK}^2$ respectively. The
errors do not significantly go down much further for $N_T <
10^{-5}{\rm mK}^2$, and we have $(0.55, 0.63, 0.63) \, \%$ at $N_T =
10^{-6}{\rm mK}^2$.

\subsection{Constraints on Neutrino mass}

 Free streaming of neutrinos causes a power suppression on large
 scales. This suppression of dark matter power spectrum shall imprint
 itself on the cross-correlation of Ly-$\alpha$ forest and 21 cm
 signal \citep{ashis1}. We have suggested this as a possible way to constrain
 neutrino mass.  We have considered a BOSS like Ly-$\alpha$ survey
 with a quasar density of $30$ deg$^{-2}$ with an average $3\sigma$
 sensitivity for the measured spectra. We have also assumed a 21 cm
 intensity mapping experiment at a fiducial redshift $z = 2.5$
 corresponding to a frequency $406$MHz using a SKA1-mid like
 instrument with 250 dishes each of diameter $15$m. We have assumed a
 $(\Omega_{\Lambda},\Omega_{m},\Omega_{r},h,\sum\nolimits_{i}m_{i})$=$(0.6825,0.3175,
 0.00005, 0.6711, 0.1 {\rm ev})$ \citep{plank13} for this analysis.
 The Fisher matrix analysis using a two parameter
 $(\Omega_m. \Omega_{\nu}) $ shows that For a 10.000 hrs radio
 observation distributed over 25 pointings of 400 hrs each the
 parameters $\Omega_m $ and $\Omega_{\nu}$ are measurable at $0.321 \%$
 and $3.671\%$.  respectively [see figure (\ref{fig:ellipse})]. We find it significant that instead of
 a deep long duration observation in one small field of view, it is
 much better if one divides the total observation time over several
 pointings and thereby increasing survey volume.  For 100 pointings
 each of $100 hrs$ one can get a $2.36 \%$ measurement of
 $\Omega_{\nu}$.  This is close to the cosmic variance limit at the
 fiducial redshift and the given observations.  In the ideal limit one
 may measure $\Omega_{\nu}$ at a $2.45 \%$ level which corresponds to
 a measurement of $\sum m_{\nu}$ at the precision of $(0.1 \pm 0.012)$ eV.
\begin{figure}
\begin{center}
  \includegraphics[height=5cm, width=5cm, angle=0]{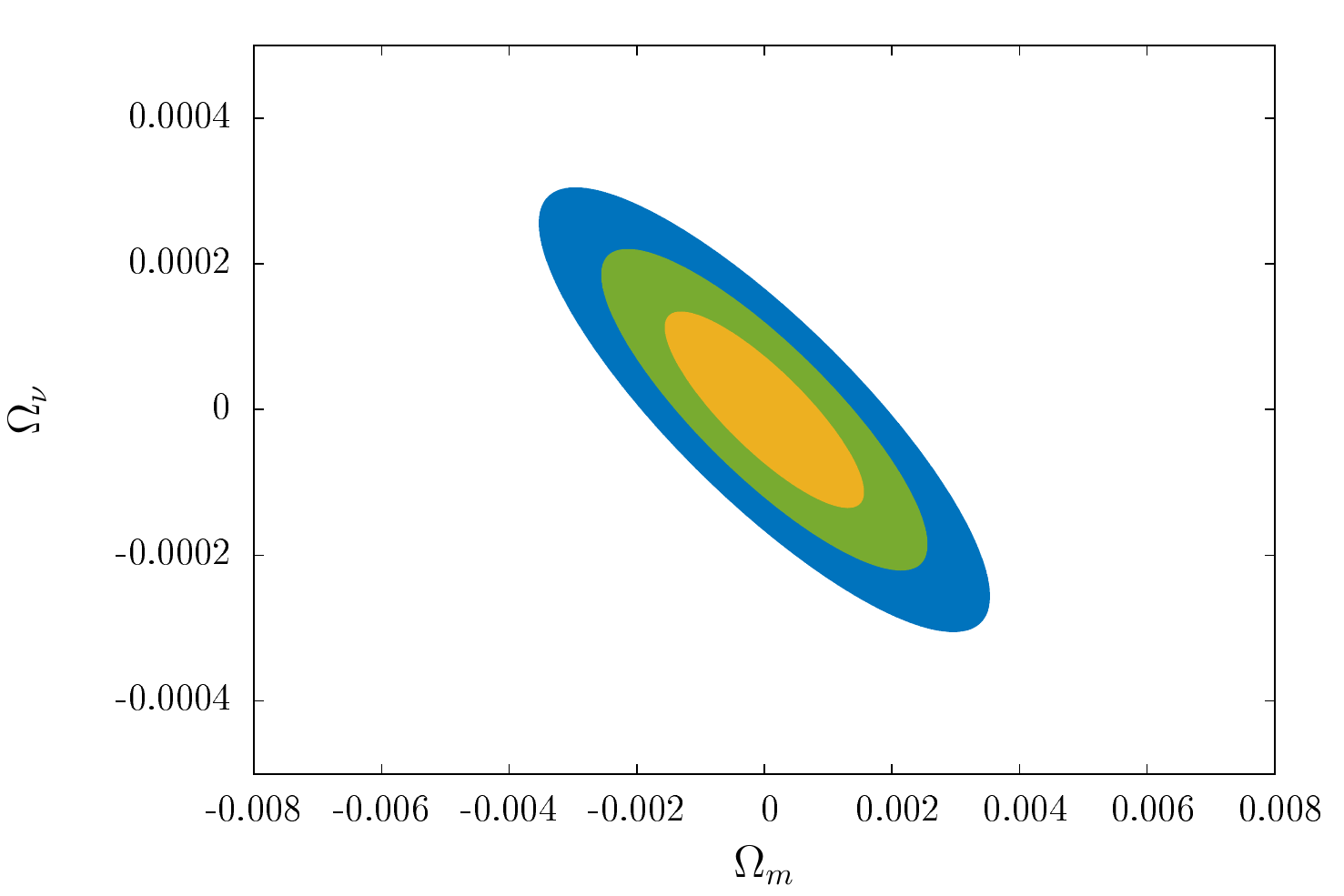}\\
  \includegraphics[height=3.5cm, width=3.5cm, angle=0]{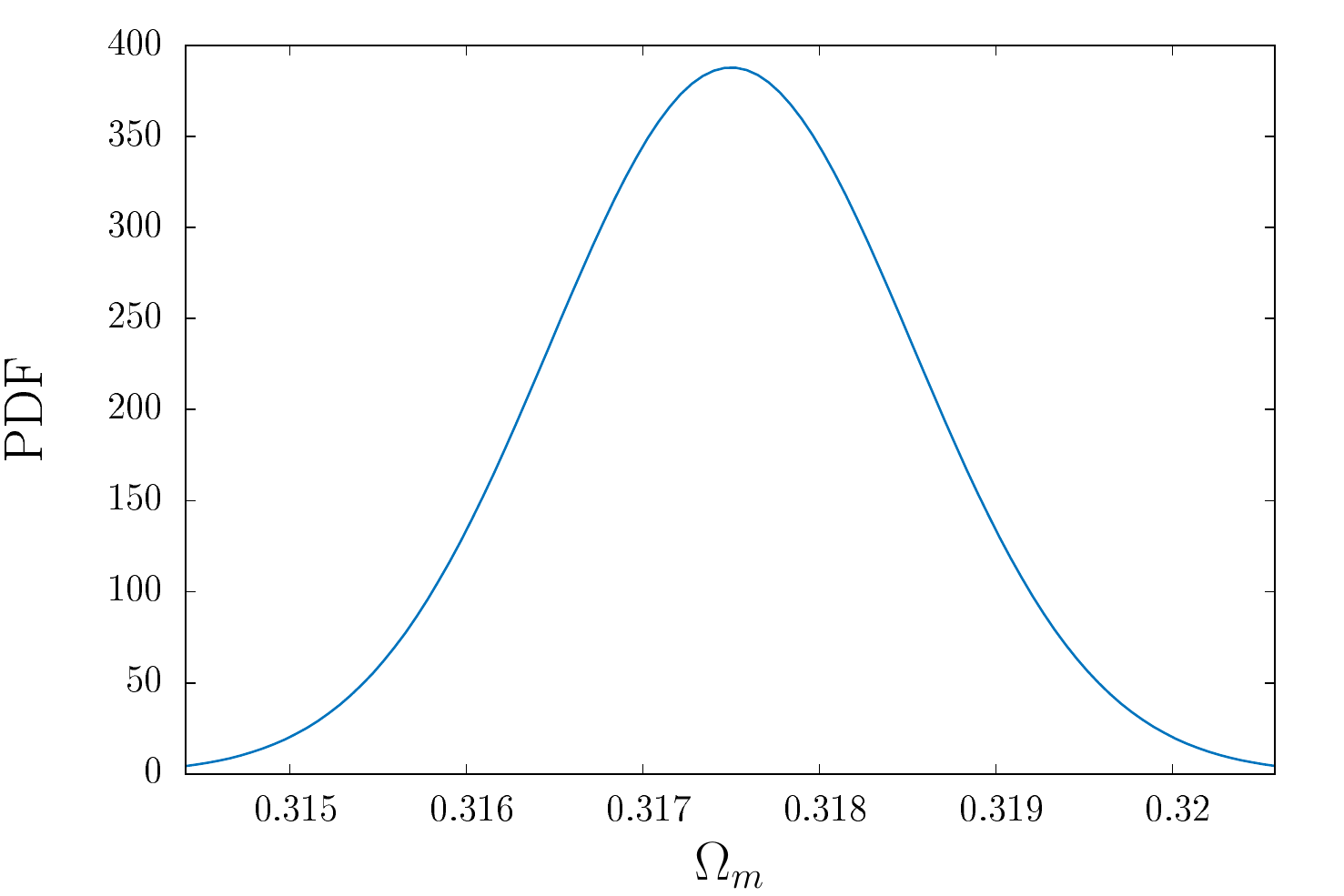}
  \includegraphics[height=3.5cm, width=3.5cm, angle=0]{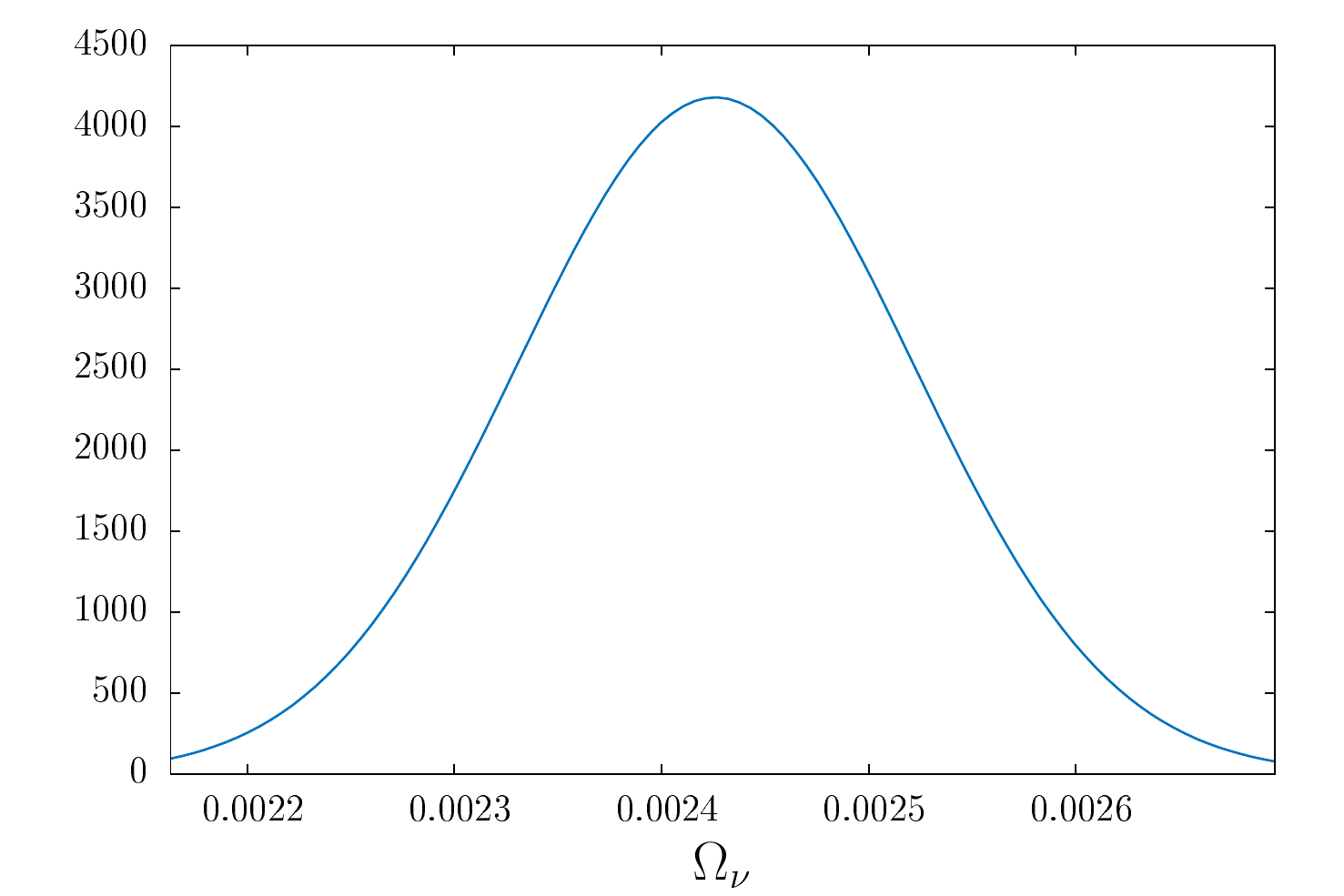}
    \caption{68.3\%,95.4\% and 99.8\% ellipse for 10000 hrs. observations for 25 pointings with each pointing of 400 hrs.  observations. The  marginalized one dimensional probability distribution function (PDF) (\cite{ashis1})
for $ \Omega_{m} $ and $ \Omega_{\nu} $ are also shown.}
\label{fig:ellipse}
\end{center}
\end{figure}
\section{Cross-correlation with Lyman break galaxies}
The cross-correlation between the HI 21-cm signal and the Lyman break galaxies is another important
tool to probe the large scale structure of the Universe at post reionization epoch. This has been studied 
recently \citep{navkkd}  using a high resolution N-body simulation.
Prospects for detecting such a signal using the SKA1-mid and SKA1-low telescopes together with a 
Lyman break galaxy spectroscopic survey with the same volume have also been investigated. 
It is seen  that the cross power spectrum can be detected with a SNR up to $\sim 10$ times 
higher than the HI 21-cm auto power spectrum. Like in all other cross power spectrum the 
Lyman break galaxy and HI 21-cm cross power spectrum is expected to be extracted more reliably
 from the much stronger by spectrally smoothed foreground contamination compared to the HI 21-cm auto
 power spectrum.

\section{Cross-correlation of HI 21 cm signal with CMBR}
\subsection{ Weak Lensing }
Gravitational lensing has the effect of  deflecting  the CMBR photons. This forms a
secondary anisotropy in the CMBR temperature anisotropy maps
(\cite{antonylenss}).  The weak lensing of CMBR  is a powerful  probe the universe at distances ($z \sim 1100$)  far greater 
than any galaxy surveys. Measurement of the secondary CMBR
anisotropies, often  uses the cross correlation of some relevant
observable (related to the CMB fluctuations) with some tracer of the
large scale structure (\cite{hirataa, smithh, padmaa}).  For weak lensing
statistics  the  `convergence' and the `shear' fields quantify the
distortion of the maps due to gravitational lensing.  Convergence ($\kappa$)
measures the lensing effect through its direct dependence on the
gravitational potential along the line of sight and is thereby a
direct probe of cosmology. The difficulty in precise measurement of
lensing is the need for very high resolutions in the CMBR maps, since typical
deflections over cosmological scales is only a few arcminutes. The
non-Gaussianity imprinted by lensing on smaller scales allows a
statistical detection for surveys with low angular resolution.
Cross-correlation with traces, limits the effect of systematics and
thereby increases the signal to noise.  The weak lensing observables
like convergence are constructed using various estimators involving the  the CMBR maps(T, E, B) (\cite{seljak-99,
  hu-apj, huokaa}). The reconstructed convergence field can then be used
for cross correlation.
        
We have probed the possibility of using the post-reionization HI
as a tracer of  large scale structure to detect the  weak
lensing \citep{tgs2} effects.   We have studied the cross correlation
between the fluctuations in the 21-cm brightness temperature maps and
the weak lensing convergence field.  We can probe the one dimensional integral effect of lensing at any intermediate redshift by
tuning the observational frequency band for 21-cm observation.  The
cross-correlation power spectrum can hence independently quantify the cosmic
evolution and structure formation at redshifts $z \le 6$. The
cross-correlation power spectrum may also be used to independently compare the
various de-lensing estimators. 
                                  
The distortions caused by the deflection is the quantity of study in
weak lensing.  At the lowest order, magnification of the signal is
contained in the convergence.  The convergence field is a line of
sight integral of the matter over density $\delta$ given by
(\cite{ludo}) \be \kappa(\n) = \frac{3}{2} \Omega_{m0} {\left(
  \frac{H_0}{c} \right)}^2 \int_{\eta_0}^{\eta_{LSS}} d\eta F(\eta)
\delta( \mathcal{D}(\eta)\n , \eta) \ee and $F(\eta)$ is given by \be
F(\eta) = \frac {\mathcal{D}(\eta_{LSS} -
  \eta)\mathcal{D}(\eta)D_{+}(\eta)}{\mathcal{D}(\eta_{LSS}) a(\eta)}
\ee Here, $D_{+}$ denotes the growing mode of  density
contrast $\delta$, and $\eta_{LSS}$ denotes the conformal time
to the epoch of recombination.  The comoving angular
diameter distance $\mathcal{D}(\chi) = \chi $ for flat universe,
$\mathcal{D}(\chi) = \sin ( K \chi) / K $ for $ K = | 1 - \Omega_m -
\Omega_{\Lambda}|^{1/2} H_0/c < 0$ and $\mathcal{D}(\chi) = \sinh ( K
\chi) / K$ for $K >0$ Universe. 
The convergence  power spectrum is defined
as $
\langle \ a_{\ell m}^{\kappa} a_{\ell' m'}^{\kappa *} \ \rangle =
C_{\ell}^{\kappa} \delta_{\ell \ell'} \delta_{m m'}$. 
where  $a_{\ell m}^{\kappa}$ are the expansion coefficients in spherical harmonic basis. The Convergence auto-correlation
power spectrum for large $\ell$ can be approximated as  
\be
\mathcal{C}_{\ell}^{\kappa} \approx \frac{9}{4}\Omega_{m0}^{2} {\left(
  \frac{H_0}{c} \right)}^4 \int d \eta\frac{ F^2(\eta) }{\mathcal{D}^2(\eta)}
P{\left(\frac{\ell}{\mathcal{D}(\eta)}\right)}
\ee
The cross correlation angular power spectrum between the post-reionization
\nh 21-cm brightness temperature signal and the convergence field, is 
given by 
\be
{\mathcal{C}}_{\ell}^{HI-\kappa} = A (z_{\rm HI}) \int dk \left[ k^2  P(k) {\mathcal{I}}_\ell(k r_{\rm HI}) \int d \eta F(\eta)
  j_{\ell}(kr)\right]
\ee 
where $P(k)$ is  dark matter power spectrum at $ z=0$ , and 
\be
 {A}(z) = \frac{3}{\pi}\Omega_{m0} {\left( \frac{H_0}{c} \right)}^2
 \bar{T}(z)\bar{x}_{\rm HI}
 D_{+}(z)    
\ee
 We note  that  the convergence field
 $\kappa(\n)$,   is
not directly measurable in CMBR experiments. It is reconstructed from the CMBR maps  through the
use of various statistical estimators (\cite{hansonn, kesdenn, cookess}). The cross-correlation angular  power spectrum, $ \mathcal{C}_{\ell}^{HI-\kappa}$, does not  de-lens the CMB
 maps directly. It  uses the reconstructed cosmic shear fields , and is thereby very sensitive to the underlying tools of  de-lensing, and the cosmological model.
The  cross-correlation angular power spectrum  may provide a
 way to independently compare various  de-lensing estimators.

The cross-correlation power spectrum follows the same shape as the matter
power spectrum. The  signal peaks at a
particular $\ell$ which  scales as $ \ell \propto  r_{\rm HI}$ when the redshift is changed. The
angular distribution of power clearly follows the underlying
clustering properties of matter.
The amplitude depends on
several  factors which are related to 
cosmological model and the \nh distribution at $
z_{\rm HI}$.
The angular diameter distances directly  also depends directly  on the cosmological parameters.
The cross-correlation signal may hence be used independently for joint estimation of cosmological parameters.

We shall now discuss the prospect of detecting the cross-correlation
signal assuming a perfect foreground removal. The error in the cross-correlation signal has 
the contribution due to instrumental noise and sample variance.
Sample variance however puts a
limiting  bound on the detectability. 
The cosmic variance for   
$\mathcal{C}_{\ell}^{HI-\kappa} $ is given by
\be
\sigma_{SV}^2 = \frac {\mathcal{C}_{\ell}^{\kappa}
 \mathcal{C}_{\ell}^{HI}}{(2\ell + 1){N_c}f_s \Delta\ell}
\ee
where  $ f_s$ is fraction  of overlap portion of sky common to
both observations. $N_c$ denotes the number of
independent estimates of the signal.

In the ideal hypothetical  possibility of a full sky 21 cm survey we have  $f_s = 1$, and
used $\Delta\ell = 1$. 
The predicted $ S/N$ is found to be $\sim 2$ and is not significantly  high
for  detection which requires $S/N
\geq 3$. Choosing a  $\Delta\ell = 10$ for $\ell \leq 100$ and
$\Delta\ell = 100$ for $ \ell > 100$ shall  produce a $S/N > 3$.
This establishes that, with full sky coverage and
negligible instrumental noise, the binned cross-correlation power
spectrum is not cosmic variance limited and it detectable. The $S/N$ estimate is based on \nh observation at
only one frequency. The $21$- cm observations allow us to probe a
continuous range of 
redshifts. This allows us to further increase the $S/N$ by collapsing
the signal from various redshifts. In principle,  a broad band 21-cm experiment may
further increase the $S/N$.

The  $S/N$ maybe improved by collapsing the signal from
different scales $\ell$ and thereby test the feasibility of a
statistically significant detection. The  cumulated SNR upto a multipole
$\ell$ is given by  
\be
{\left(\frac{S}{N} \right)}^2 = \sum_{0}^{\ell} \frac{(2\ell' + 1) {N_c}
  f_s{(\mathcal{C}_{\ell'}^{HI-\kappa})}^2
}{(\mathcal{C}_{\ell'}^{HI} + N_{\ell'}^{HI}
  )(\mathcal{C}_{\ell'}^{\kappa} + N_{\ell'}^{\kappa}) }
\ee
$ N_{\ell}^{\kappa}$ and $ N_{\ell}^{HI}$ denotes the noise power
spectrum for $\kappa$ and \nh observations respectively.
Ignoring the instrument noises we note  that there is a significant increase in the $S/N$ by cumulating over multipoles $\ell$.
This implies that a statistically significant detection of
$\mathcal{C}_{\ell}^{HI-\kappa} $ is possible and the signal is not limited cosmic variance.
It is important to push instrumental noise to the limit set by cosmic variance for a detection of the signal. At the relevant redshifts of interest, it is possible to reach such low noise levels with SKA. It is however important to scan large parts of the sky and thereby increase the survey volume.

Instrumental noise plays an important role at large multipoles (small
scale).
For a typical CMB experiment,  the noise power spectrum  (\cite{marr,
kaplingg})  is given by
$
N_{\ell} = \sigma^2_{\rm{pix}}\Omega_{\rm{pix}}{ W_{\ell}}^{-2}
$, where different pixels  have uncorrelated noise with
variance $\sigma^2_{\rm{pix}}= s^2/ t _{\rm{pix}} $. Here $ s^2$
and $t_{\rm{pix}}$  are  the  pixel sensitivity and `time spent on the
pixel' respectively. $\Omega_{\rm{pix}} $ is the solid angle subtended
per pixel and we use  a Gaussian beam $ W_{\ell} = \rm{exp} [ { - \ell^2
  \theta^2_{FWHM}/ 16  ln 2}] $.

For \nh observations, the quantity of interest is the complex Visibility
which is used to estimate the power spectrum (\cite{fg3}).
For a radio telescope   with N antennae,  system temperature  $
T_{sys}$,  operating at a  frequency
$\nu$, and  band width $B$ the  noise correlation is given by
$N_{\ell}^{HI} \propto  \frac{1} {N ( N - 1)}{ \left[
    \frac{T_{sys}}{K} \right ]}^2 \frac{1}{T \sqrt{{\Delta \nu}
    B}}$.

Where  $T$ denotes total observation time, and
    $K$ is related to the effective collecting area of the antenna
    dish. Binning in $\ell$ also reduces the noise.
The bin $\Delta \ell  = 1/ 2 {\pi}^2 \theta_0$ is chosen assuming a Gaussian
    beam of width $\theta_0$.
With a  SKA like instrument (\cite{fg3}),
one can attain a noise level much lesser than the
signal by increasing the observation time (infact a 5000 hour observation
     with  present SKA configuration is good enough)  and also by increasing
the radial distance probed bi increasing the  band width of the telescope. Being  inversely related
to the total number of antennae in the radio array,  future designs may actually allow
further reduction  of the the system noise and achieve $ N_{\ell}^{HI} << \mathcal{C}_{\ell}^{HI}$.
This  establishes  the detectability of the cross-correlation signal.
We would like to conclude by noting that correlation between weak lensing fields and 21 cm maps, 
 quantified through $  \mathcal{C}_{\ell}^{{HI}-\kappa}$ may allow an independent means to estimate
 cosmological parameters and also test various estimators for CMBR delensing.

\subsection{ ISW effect}
In an Universe, dominated by the cosmological constant, $\Lambda$ , the expansion factor of the universe, $a$,
grows at a faster rate than the linear growth of density perturbations. This  consequently implies that, the 
 gravitational potential $ \Phi \propto - \delta/a$  will  decay. 
The ISW effect is caused
by the change in energy of CMB photons as they traverse
these  time dependent potentials.

If the horizon size at the  epoch of dark energy dominance (decay of
the potential) is represented by $\eta_{\Lambda}$,  then  the ISW effect is
suppressed on scales $ k  \geq 2\pi / \eta_{\Lambda}$. This
corresponds to an angular scale $\ell_{\Lambda} = 2\pi d
/ \eta_{\Lambda}$, where $d$ is the angular diameter distance to the
epoch of decay.

The ISW term in CMBR temperature anisotropy is given by
\be
 \Delta T(\n)^{\rm ISW} = 2 T \int_{\eta_{\rm{LSS}}}^{\eta_{0}} d\eta
 \eta \, \dot{\Phi}(r \n,\eta). 
\ee
The cross correlation angular power spectrum between HI 21 cm signal and ISW is given by \citep{tgs1}
\be
{\cal{C}}_{\ell}^{HI-ISW} = {\cal{K}}(z_{\rm HI}) \int dk ( P(k) \cal{I}_{\ell}(k r_{\rm HI}) \int_{\eta_{\rm LSS}}^{\eta_0} d \eta F(\eta) j_{\ell}(kr)  )
\label{eq:cl1}
\ee
where $P(k)$ is the present day dark matter power spectrum, 
\be
 \mathcal{K}(z) = -\bar{T}(z)\bar{x}_{\rm HI}
 D_{+}(z)  \frac{6 H_{0}^3 \Omega_{m0}}{\pi c^3}  
\ee
\be
{\mathcal{I}}_{\ell}(x) = b j_{\ell}(x)- f \frac{d^2 j_{\ell}}{dx^2} 
\ee
and 
\be
F(\eta) =  \frac{D_{+} (f-1) H(z)}{H_0}
\ee
For large $\ell$ we can use the Limber approximation (\cite{lim, Afshordi44})
which allows us to replace the spherical Bessel functions by a  Dirac
deltas as \[ j_{\ell}(kr) \approx  \sqrt{\frac{\pi}{2\ell+1}} \delta_D(\ell + \frac{1}{2} - kr) \]
whereby  the angular cross-correlation power spectrum 
takes the simple scaling of the form 
\be   \mathcal{C}_{\ell}^{HI-ISW}  \sim  \frac{\pi  \mathcal{K} F} {2\ell^2}
P(\frac{\ell}{r})
\label{eq:fnll}
\ee
where $P(k)$ is the present day dark matter power spectrum and all the
other terms on the {\it rhs.} are evaluated at 
$z_{\rm HI}$.  
The dimensionless quantity $f$ quantifies the growth of the dark matter 
perturbations, and
the ISW effect is proportional to $f-1$. The term $f-1$ is a sensitive
probe of  dark energy. 
Here we estimate  the viability of detecting the \nh-ISW
cross-correlation signal. 
 Cosmic variance sets a
limit  in deciding whether the signal can  
at all be detected or not. 
Even in the cosmic variance limit at $z \sim 1.0$ with a 32 MHz observation  we find that ${\rm S/N} <
0.45$ for  all $z_{\rm HI}$ and $\ell$ and a statistically significant
detection is not possible in such cases. It is possible to
increase  ${\rm S/N}$ collapsing  the signal at different multipoles
$\ell$.   To test if a statistically significant detection is thus
feasible we have collapsed all multipoles less than $\ell$ to evaluate
the cumulative ${\rm S/N}$ defined as (\cite{cooray2,crosshi33})
We find that the contribution in the cumulated $S/N$
comes from $\ell<50$ at all redshifts $ 0.4 < z < 2 $. The
cross-correlation signal is largest at ($z \sim 0.4$) and is
negligible for ($ z > 3$). We further find that although there is an
increase in ${\rm S/N}$ on collapsing the multipoles it is still less
than unity. This implies that a statistically significant detection is
still not possible. Thus, probing a thin shell of \nh doesn't allow us
to detect a cross correlation, the signal being limited by the cosmic
variance.  A cumulated S/N of $ \sim 1.6 $ is attained for redshift
upto $z = 2$ and there is hardly any increase in S/N on cumulating
beyond this redshift.  This is reasonable because the contribution
from the ISW effect becomes smaller beyond the redshift $z>2$. This
S/N is the theoretically calculated value for an ideal situation and
is unattainable for most practical purposes. Incomplete sky coverage,
and foreground removal issues would actually reduce the S/N and
attaining a statistically significant level is not feasible.

\section{Acknowledgements}  KKD would like to thank DST for support through the project SR/FTP/PS-119/2012.
TGS would like to thank the  project SR/FTP/PS-172/2012 for financial support.

\label{lastpage}

\end{document}